# Effective Potential Energy Expression for Membrane Transport

ROBERT W. FINKEL
Physics Department
St. John's University
Jamaica New York 11439
USA
finkelrt@stjohns.edu

*Abstract:* - All living cells transport molecules and ions across membranes, often against concentration gradients. This active transport requires continual energy expenditure and is clearly a non-equilibrium process for which standard equilibrium thermodynamics is not rigorously applicable. Here we derive a non-equilibrium effective potential that evaluates the per-particle transport energy invested by the membrane. A novel method is used whereby a Hamiltonian function is constructed using particle concentrations as generalized coordinates. The associated generalized momenta are simply related to the individual particle energy from which we identify the effective potential. Examples are given and the formalism is compared with the equilibrium Gibb's free energy.

## 1 Introduction

Molecules and ions are transported across cellular membranes by various means; these are generally classified as diffusion, facilitated diffusion, and active transport. Except for simple diffusion, the membranes can supply energy to the process. Membrane energy may, for example, be used to power transmembrane enzymes that pump ions and molecules from one side to the other [1]. Although standard thermodynamic relations are regularly and successfully used to treat transport processes, such relations are strictly valid only at equilibrium. However, molecular and ionic flows that continually expend energy are non-equilibrium processes and it is desirable to formulate energy relations that work for transport systems operating away from equilibrium.

The approach taken here is to first write rate equations that describe concentration flows across the membrane. These are chemical kinetic equations for concentrations on either side of a membrane and they express the flow rates due to both diffusion and active or facilitated transport. We then enlist a formalism to construct a Hamiltonian function from the rate equations. The generalized coordinates for this Hamiltonian are the chemical concentrations and the corresponding generalized momenta are related to the average individual particle energies. Upon calculating the per-particle energy, the result appears as the sum of a chemical potential and an effective potential. The effective potential energy is then identified with the energy that the membrane invests in the per-particle transport.

## 2 Hamiltonian Formulation

We consider particle transport across a membrane by some energetic agent. Let the lower and higher relative concentrations on either side of the membrane be represented by $n_1$ and $n_2$, respectively. Two competing processes then determine the flow rate, diffusion tending to increase the less dense solute and the opposing agent of active transport carrying solute in the opposite direction. The associated linear chemical kinetic equation is

$$\dot{n}_1 = a(n_2 - n_1) - bn_1 \qquad (1)$$

where the dot notation indicates differentiation with respect to time, coefficient $a$ is associated with Fick diffusion [2], and coefficient $b$ summarizes the transport effect of an energetic agent. A similar equation for $n_2$ follows from the requirement for relative concentrations,

$$n_1 + n_2 = 1. \qquad (2)$$

Equations (1) and (2) are readily solved, but here we are interested in the energy of transport. Toward that end, we treat the $n$'s as generalized coordinates and incorporate the rates into a known Hamiltonian form, $H$ [3],

$$H = \sum_i (\dot{n}_i p_i + n_i \varepsilon_i), \qquad (3)$$

where the $\varepsilon$'s are "rest energies" and the $p$'s are generalized momenta. The rest energies consist of

membrane potential energies $k_B T U_m$ and ideal chemical potentials, $\mu_i = k_B T \ln(n_i) + \mu_i^0$

$$\varepsilon_i = k_B T (\ln n_i + U_m) + \mu_i^0 \qquad (4)$$

where $k_B$ is Boltzmann's constant, $T$ is Kelvin temperature, and $\mu_i^0$ is a standard potential. We arbitrarily choose the zero of the membrane potential to be on the high concentration side of the membrane. The generalized momenta will provide the energies, $E_i$, of average individual particles according to the relation [4],

$$E_i = -\dot{p}_i \qquad (5)$$

## 3 General Solution

We focus on finding an expression for the average per-particle energy $E_1$. Apply Hamilton's equation [5], $\dot{p}_1 = -\partial H / \partial n_i$ and from equation (5) we have

$$E_i = k_B T \ln n_1 - (a+b)P + k_B T + \mu_1^0 \qquad (6)$$

with

$$P \equiv p_1 - p_2.$$

Now $P$ must be evaluated in order to turn (6) into a useful form.

The Hamiltonian for an autonomous system is some constant, $E$, so we set $H = E$ and solve for $P$. However, the denominator of the resulting expression vanishes at the steady state $\dot{n}_1 = 0$. To assure a finite result for $E_1$, choose the undetermined constant $E$ so that the numerator of $P$ also vanishes at the steady state. Let superscripts "0" represent values at the steady state so that $E$ can be written as

$$E = n_1^0 \varepsilon_1^0 + n_2^0 \varepsilon_2^0. \qquad (7)$$

This choice will produce a finite limit for $P$ at the steady state.

The central expression for per-particle energy can now be completed. Substituting $P$ into equation (6) gives our sought-for expression for $E_1$,

$$E_1 = \mu_1 + U + k_B T, \qquad (8)$$

where $U$ is an effective potential that summarizes the energy attributable to the non-equilibrium process,

$$U = \frac{k_B T (n_1 \ln n_1 + n_2 \ln n_2 + n_1 U_m) - E}{1 - (r+1)n_1} \qquad (9)$$

where $r$ is the ratio of equilibrium concentrations,

$$r = n_1^0 / n_2^0. \qquad (10)$$

Note that $r$ is the only free parameter in equation (8) because the steady state concentrations are expressed in terms of $r$:

$$n_1^0 = r/(1+r), \quad n_2^0 = 1/(1+r) \qquad (11)$$

The isolated term $k_B T$ in equation (8) is associated with every particle [6] and is disregarded here because only differences in energy can contribute to dynamics.

## 4 Steady State Solution

Although equations (8) and (9) apply to general non-equilibrium cases, most measurements take place at the steady state. There the form for $U$ is particularly simple. Take the limit of $U$ as $n_1 \to n_1^0$ to obtain

$$U \to -\frac{k_B T (\ln r + U_m)}{r+1}. \qquad (12)$$

$U$ is formulated from the viewpoint of the lower concentration solution and we arbitrarily ascribe the higher concentration regions to be at zero potential.

Some features of (12) are worth noting. At an experimental temperature of $T = 296$ K and membrane potential of ~70 mV, the effective potential is positive when $r < \frac{1}{15}$ and negative when $r > \frac{1}{15}$. The sodium-potassium pump illustrates this; extracellular potassium has $r \approx \frac{1}{20}$ giving $U$ a value of ~0.24 mV so the membrane adds negligible energy to transport $K^+$ ions out of the cell. (That is, $K^+$ transfer is close to spontaneous.) Conversely, the lower concentration of sodium is on the intracellular side of the cell membrane and has a relative concentration of $r \approx \frac{1}{10}$. From this side the membrane potential is seen as –70 mV. The effective potential then shows that the membrane actively exerts 117 mV of energy per $Na^+$ ion transferred or, equivalently, 2.7 kcal per mole. The sodium/potassium pump moves three Na+ ions in each cycle requiring 8 kcal/mole in agreement with experiment [7].

## 5  Discussion

The energy relations presented here have the benefit of being generic and simple. They also assure more accurate membrane energies than equilibrium calculations. For example, compare the steady state equation (12) with the standard equilibrium expression for the Gibb's free energy. In our notation the change in Gibb's free energy per particle is $-k_B T \ln r$. We see that the accuracy of the equilibrium form necessitates that $r$ is sufficiently small.

This note also extends energy analyses to transient transfers. The effective potential of equation (9) is not limited to the steady state; we anticipate that in future investigations this may help to analyze transient absorption of medications or nutrients through the plasma membrane.